\documentstyle[preprint,aps]{revtex}

\begin{document}
\draft
\preprint{IMSc/2000/09/52}
\title{Notes on Isolated Horizons}
\author{G. Date \footnote{e-mail: shyam@imsc.ernet.in}}
\address{The Institute of Mathematical Sciences,
CIT Campus, Chennai-600 113, INDIA.\\}

\maketitle
\begin{abstract} 

A general analysis for characterizing and classifying `isolated horizons' 
is presented in terms of null tetrads and spin coefficients. The freely
specifiable spin coefficients corresponding to isolated horizons are
identified and specific symmetry classes are enumerated. For isolated
horizons admitting at least one spatial isometry, a standard set of
spherical coordinates are introduced and associated metric is obtained.
An angular momentum is also defined.  

\end{abstract}

\vskip 2.0cm

\pacs{PACS numbers:  04.20.-q, 04.70.-s } 

\narrowtext

\section{Introduction}

Stationary black holes in asymptotically flat space-time obey the famous
laws of black hole mechanics \cite{stationary} which admit a
thermodynamical interpretation\cite{history} supported by the Hawking effect
\cite{hawking}. These provide an arena to study the interplay of classical 
gravity, quantum mechanics and thermodynamics \cite{interplay}. These involve 
the global notions of event horizon, asymptotic flatness and stationarity. 
There have been two generalizations by relaxing some of the global
conditions, namely the `trapping horizons' of Hayward \cite{trapping} and 
the `isolated horizons' of Ashtekar et al \cite{isolated}. \\

Hayward \cite{trapping} introduced the quasi-local idea of ``trapping" 
horizons by abstracting the property that these are foliations of
(suitably) marginally trapped surfaces. He gave a classification of ``trapping
horizons" and also defined quantities satisfying a generalized version
of all the laws of black hole mechanics. The entire analysis is strictly
quasi-local with {\it{no}} reference to any asymptotics. \\

Among these classes of horizons, is a special case wherein a horizon is a null 
hyper-surface. These horizons have the property that the area of its two 
dimensional space-like surfaces remain constant along its null generators 
and the second law reduces trivially to the statement that the area is constant.
Since one expects these areas to change only when there is energy flow across 
the horizon, these horizons suggests an intuitive idea of `isolation'. \\ 

With a somewhat different notion of ``isolated horizons", recent work of 
Ashtekar et al \cite{isolated} also seeks to replace event horizons and yet get
analogue of the zeroth and the first law of black hole mechanics. In
this generalization, though reference to asymptotics is retained,
stationarity of usual black holes is relaxed. In particular, Ashtekar et al 
formulate these notions and their associated quantities such as mass, in 
terms of variables and actions suitable for passage to a non-perturbative 
quantization. This generalization permits one to deal with a situation 
wherein a collapse proceeds in stages punctuated by a series of `non activity'
across a collapsed `core'. Each of such stage of `non activity' is modeled by 
an ``isolated horizon". Loss of stationarity come with a price which makes 
this generalization non trivial and leads to infinitely many forms of the 
first law \cite{distorted}. This framework allows  Ashtekar et al to compute 
the entropy of the so called non-rotating isolated horizons \cite{entropy}. \\

The latter notion of isolated horizons is more restrictive than that of 
trapping horizons in the sense that isolated horizons are null hyper-surfaces 
while trapping horizons have no such restriction. The restriction is
not overly strong, the space of solutions admitting isolated horizons is
infinite dimensional \cite{lewandowski}. The restriction however gives
some control over the space of solutions and recently Ashtekar et al
\cite{distorted} have given a very interesting interpretation of the first 
law(s). \\

In both the generalizations mentioned above, there are two part. The
first is the geometric characterization of appropriate horizon. This
involves both the distinguishing features of appropriate three dimensional 
sub-manifold of a space-time and the definitions of quantities such as surface
gravity, mass, area, angular velocity, angular momentum, charge etc associated 
with the horizon. The second part involves showing that the quantities so
defined do satisfy the laws of black hole mechanics (or a generalized version 
thereof). \\

In this work we consider a uniform treatment of general isolated horizons 
only. Our aim is to arrive at a characterization such that the zeroth and 
the first law of black hole mechanics hold. One should observe at the out 
set that the zeroth law is a property of a {\em {single}} solution of 
matter-Einstein field equation admitting an isolated horizon. The first law 
however involves comparison of certain quantities associated with several 
such solutions and thus is a statement about properties of the full space 
of such solutions. Establishing the first law is therefore expected to be 
more involved. \\

A remark about the slight difference in the approach of Ashtekar and
co-workers and ours is in order. In the approach of Ashtekar et al, 
an action principle plays an important role, particularly for the first
law and of course in the subsequent computation of the entropy. It is
thus natural to view an isolated horizon as an ``inner boundary" of a
suitable class of space-time manifolds and transcribe characterization
of isolated horizons in terms of boundary conditions on the variables
entering the action formulation. On the other hand, as a first step, we
wish to translate the physical idea behind isolated horizon in to a geometrical
characterization and try to get a handle on the space of solutions
admitting such horizons. For this an action formulation is unnecessary
and it is natural to view an isolated horizon as a suitable null
hypersurface of a solution space-time. Since a tetrad formulation is closer to
(metric) geometry, it is a natural choice for us. This has the further
advantage that it can be generalized to other dimensions. If one could
get the laws of black hole mechanics without the use of an action
principle, then one could also treat phenomenological matter. One should
also remark that since both approaches capture the same physical idea,
one does not expect different results at the level of geometrical
characterization. What we do seek though is a direct and systematic
arrival at a geometrical characterization of the physical notions.\\

The basic defining property of these horizons is that they are all
null hyper-surfaces. This naturally suggests the use of null tetrad
formulation. In terms of suitably adapted null tetrads and their
corresponding spin coefficients, this alone immediately implies that
the coefficients $\kappa = 0$ and $\rho$ is real. 
To this one adds the requirement
that the expansion of the null geodesic generators be zero ($\rho = 0$)
which incorporates
the properties of being isolated (constancy of area) and potential
marginal trapping. Since these are supposed to be hyper-surfaces of 
physical space times, it is only natural to require that Einstein equations 
hold and that the stress tensor satisfies a suitable energy condition. 
Raychoudhuri equation and the energy conditions then imply a number of 
consequences, one of which is that the null geodesic congruence is also 
shear free ($\sigma = 0$). \\

There is however a good deal of freedom in choosing the null tetrads,
the freedom to make local Lorentz transformations. These have been
conveniently classified into three types \cite{chandra}. Having gotten a 
null direction field, the relevant freedom is reduced to only the so-called 
type-I and type-III transformations. Due to this freedom, one can not 
characterize isolated horizons in terms of the spin coefficients by themselves 
unless these are invariant under the local Lorentz transformation. 
(The particular values of the spin coefficients in the previous paragraph are 
indeed invariant.) One can either impose conditions on quantities that are 
invariant under these local Lorentz transformations or one could fix a 
convenient `gauge' and then use the corresponding spin coefficients for 
characterization. In this work we use a combination of both.\\

The paper is naturally divided into two parts. The first part uses a
gauge fixing procedure to identify freely specifiable spin coefficients.
The second part uses invariant quantities to analyze symmetry classes of 
the horizons and identifies suitable associated quantities. \\ 

The paper is organized as follows. \\

In section II we discuss the basic conditions characterizing isolated
horizons and arrive at the zeroth law. We also obtain the `freely
specifiable' spin coefficients by a gauge fixing procedure. \\

In section III we discuss a symmetry classification (isometries). 
A characterization of ``rotating horizons" is discussed and an angular
momentum is identified. This section uses invariant quantities and
constructs a natural set of coordinates on the null hyper-surface. We
assume the existence of at least one spatial symmetry for this purpose.
A candidate parameterization of a `mass' is also given.\\

Section IV contains a summary and remarks on the first law. \\

The notation and conventions used are those of Chandrasekhar
\cite{chandra} and some of these are collected in appendix A for
reader's convenience. The metric signature is \\
(+ - - -). \\

Appendix B is included to illustrate our procedures for the Kerr-Newman
family. \\

\section{Characterization of Isolated Horizons}
\subsection{Basic Conditions and the Zeroth Law}

Our definition of isolated horizon will turn out to be the same as the
most recent one given by Ashtekar et al \cite{distorted}. We will
however {\it{ not}} try to define it completely intrinsically nor view
it as a boundary to be attached to an exterior asymptotically flat 
space-time. We will explicitly view an isolated horizon as a null 
hyper-surface in a four dimensional space-time which is a solution of a set
of Einstein-matter field equations. The matter stress tensor is required
to satisfy suitable energy condition and the solution is required to be
causally well behaved. To keep the logic and the role of each of the conditions 
transparent, we will begin with minimal conditions, see their implications 
and add further conditions as needed. \\

{\bf{(I)}} An isolated horizon is a null hyper-surface in a solution of
four dimensional Einstein-matter field equations with matter satisfying
the dominant energy condition (which implies also the weak energy and
the null energy conditions). Thus, Einstein equation holds on $\Delta$ and 
the matter stress tensor satisfies: 

\begin{itemize}
\item The weak energy condition and in particular, 
$T_{\mu\nu}\ell^{\mu}\ell^{\nu} \ge 0$; 

\item $T_{\mu\nu}\ell^{\nu}$ is causal. It is future(past) directed according
as $\ell^{\mu}$ is. 

\end{itemize}

One can always write these equations with reference to a chosen tetrad. 
The null hyper-surface character of $\Delta$, singles out one null direction 
and makes the use of null tetrad natural. We will thus assume that we have 
made a (arbitrary) choice of null tetrad, $\ell, n, m, \bar{m}$, such that 
$\ell$, at points of $\Delta$, is along the direction of the null normal. 
There is of course a large class of null tetrads to choose from and this is 
parameterized by the group of type-I and type-III local Lorentz transformations 
(see appendix A). We will refer to the type-I transformations as {\it {(complex)
 boosts}}, type-III scaling as {\it{scaling}} and type-III rotations of $m, 
\bar{m}$ as {\it{rotations}}.\\

The hyper-surface orthogonality of $\Delta$ implies that the null
congruence defined by $\ell^{\mu}\partial_{\mu}$ is a geodetic congruence
($\kappa$ = 0) and is twist-free ($\rho$ is real). The geodesics however
are not affinely parameterized in general. The shear of this congruence
is given in terms of $\sigma$. Vanishing of $\kappa$ already  makes the
twist and shear to be invariant under of boost transformations. Furthermore
the null geodesics generate $\Delta$ and hence $\Delta$ has the topology
of $R \times \Sigma_2$ (We assume there are no closed causal curves). 
The $R$ part can be an interval while $\Sigma_2$ could be compact or 
non-compact. At present, one need not stipulate these global aspects. But
later we will restrict to $\Sigma_2$ being spherical. The topology, 
guarantees existence of at least one foliation, not necessarily 
unique, as follows. \\


Choose any two dimensional sub-manifold $\Sigma_2$ of $\Delta$ such that
it is transversal to the direction field of the null normals. Fix an arbitrary 
null vector field along the null direction. Under diffeomorphisms generated 
by the null vector field, we will generate images of $\Sigma_2$ and hence a 
foliation of $\Delta$. The leaves of such a foliation will be diffeomorphic 
to $\Sigma_2$. One will therefore have a natural (and arbitrary) choice of 
null directions tangential to the leaves. Once we make a choice of $m, 
\bar{m}$, the null tetrad is uniquely completed. It then follows immediately 
that the null congruence defined by $n^{\mu}\partial_{\mu}$ is also twist-free 
($\mu$ is real). Since the leaves are obtained by diffeomorphism generated 
by $\ell$ and the $m, \bar{m}$ are tangential to the leaves, it follows that ,

\begin{equation}
({\cal{L}}_{\ell}m)^{\mu} n_{\mu} = ({\cal{L}}_{\ell}\bar{m})^{\mu} n_{\mu} = 0 
\end{equation}

This in turn implies that $(\alpha + \bar{\beta} - \pi) = 0 $ must
hold. Note that this procedure can always be followed and hence we can
always have a choice of null tetrad such that $\mu$ is real and $\alpha
+ \bar{\beta} = \pi$. \\

{\underline {Remark:}} In the above we made two arbitrary choices, the
vector field $\ell$ and the initial transversal $\Sigma_2$. We have {\it{not}}
defined $m, \bar{m}$ by any transport from those chosen on initial $\Sigma_2$.
There is no loss of generality though. By the available freedom of making 
local boosts, scaling and rotations, we can change the initial null normal by 
local scaling and also the initial leaf (and hence the foliation) by local 
boosts. These transformations are not completely general though since we want 
to preserve the foliation property. We will refer to these restricted
transformations as residual transformation. These will get progressively
further restricted. At this stage, the rotation parameter is completely free 
and so is the scaling parameter while the boosts parameter satisfies,

\begin{eqnarray}
D a - (\epsilon - \bar{\epsilon}) a & = &  0  \nonumber \\
D \bar{a} + (\epsilon - \bar{\epsilon}) \bar{a} & = &  0  \\
\delta \bar{a} - (\bar{\alpha} - \beta) \bar{a} & = &  
	\bar{\delta} a - (\alpha - \bar{\beta}) a \nonumber
\end{eqnarray}

Thus from the null hyper-surface property and a procedure of choosing
null tetrads we have deduced that for every given initial
choice of null normal vector field and initial $\Sigma_2$, there exist a
choice of null tetrads such that the spin coefficients satisfy, $\kappa = 
\rho - \bar{\rho} = \mu - \bar{\mu} = \alpha + \bar{\beta} - \pi = 0$. 
Further more we are free to make the residual transformations. \\

For the next condition one can give two different arguments. The $\ell$ and 
the $n$ congruences are orthogonal to the leaves. If the $n$ congruence were 
also geodesic ($\nu = 0$) then depending on the expansion of the 
$n$-congruence, we could have the leaves as (marginally) trapped surfaces. 
Indeed we would like to have this property, at least in the black hole
context, so at the least {\it{we require $\ell$ to be expansions-free }}
($\rho + \bar{\rho} = 0$). We will however leave $\nu $ unspecified for the 
moment. \\

Alternatively, this condition also shows that the induced metric on the
leaves does not vary from leaf to leaf and hence the area of a leaf is a
constant or equivalently, the Lie derivative of $m \wedge \bar{m}$
projected to $\Sigma_2$ is zero. Since the area is expected to change
only when there is energy flow across the horizon, this captures the
notion of ``isolation". In the terminology of Ashtekar et al
\cite{distorted}, this is specification of {\it{non-expanding horizon}}. \\

{\bf{(II)}} $\ell$ congruence is expansion-free ($\rho + \bar{\rho} = 0$). \\

A number of consequences follow from these conditions. The first of the energy 
conditions in conjunction with the Raychoudhuri equation for the expansion 
of $\ell$ congruence and the zero expansion condition above implies that 

\begin{itemize}
\item shear($\ell) = 0$ ($\sigma = 0$), 
\item $R_{\mu\nu}\ell^{\mu}\ell^{\nu} = 0$ on $\Delta$ and by Einstein 
equations 
\item $T_{\mu\nu}\ell^{\mu}\ell^{\nu} = 0$ on $\Delta$. 
\end{itemize}

Note that since the $\ell$ geodesics are {\it{not}} affinely parameterized, the 
Raychoudhuri equation has an extra term $(\epsilon + \bar{\epsilon})\theta
(\ell)$. This however does not affect the conclusion. \\

The $T^{\mu}_{\nu}\ell^{\nu}$ being causal and $T_{\mu\nu}\ell^{\mu}\ell^{\nu} = 
0$ then implies that $T^{\mu}_{\nu}\ell^{\nu} = e \ell^{\mu}$, with
$e$ being non-negative. Again using Einstein equation this implies that 

\begin{equation}
R_{\mu\nu}\ell^{\nu} = (8\pi e + \frac{R}{2} ) \ell_{\mu}
\end{equation}

Hence we get the Ricci scalars $\Phi_{00} = \Phi_{01} = \Phi_{10} = 0$.

Furthermore, using definitions of the Ricci scalars we deduce,

\begin{equation}
\Phi_{11} + 3 \Lambda = -4 \pi e ~~~~~~~ \equiv - {\cal{E}} ~~~~~~( \le 0).
\end{equation}

For future use we also note that the conservation of the stress tensor
and the above form for it implies that,

\begin{equation}
\ell \cdot \nabla {\cal{E}} \equiv D {\cal{E}} = 0 ;
\end{equation}

The Bianchi identities from appendix A imply,

\begin{equation}
D (\Psi_2 - \Phi_{11} - \Lambda ) = 0
\end{equation}

The equation (h) from item 5 of appendix A gives, 

\begin{equation}
\Psi_2 + 2 \Lambda = - \delta \pi - \pi (\bar{\pi} -\bar{\alpha} +
\beta) + D \mu + \mu (\epsilon + \bar{\epsilon}) 
\end{equation}

Combining the above equations we get,

\begin{equation}
-K \equiv \Psi_2 - \Phi_{11} - \Lambda = - \pi \bar{\pi} -\{ \delta \pi -
(\bar{\alpha} - \beta) \pi \} + D \mu + \mu (\epsilon + \bar{\epsilon}) +
 {\cal{E}} .
\end{equation}

And $D$ of the L.H.S. is zero by the Bianchi identity. The $K$
introduced above is the (complex) curvature of $\Sigma_2$ as defined by
\cite{penrose}. {\it{Restricting now to compact leaves without boundaries}} 
one has,

\begin{eqnarray}
\int_{\Sigma_2} ( K + \bar{K} ) & = & 4 \pi (1 - genus)  ~~~~~ (Gauss-Bonnet) \\
\int_{\Sigma_2} ( K - \bar{K} ) & = & 0 ~~~~~~~~~~~~~~~~~~~~~~ (Penrose)
\end{eqnarray}

We note in passing that the following can be checked explicitly: 

\begin{eqnarray}
\omega_+ \equiv \pi m + \bar{\pi} \bar{m} ~~~ \Longrightarrow ~~~
d \omega_+ \sim (\underline{\delta}\pi -
\underline{\bar{\delta}}\bar{\pi}) m \wedge \bar{m} \nonumber \\
\omega_- \equiv i ( \pi m - \bar{\pi} \bar{m} ) ~~~ \Longrightarrow
~~~  
d \omega_- \sim (\underline{\delta}\pi +
\underline{\bar{\delta}}\bar{\pi}) m \wedge \bar{m} 
\end{eqnarray}

Here the underlined derivatives are the compacted (rotation covariant)
derivatives defined in appendix A. From this the equation (10) follows and 
integral of the right hand side of the second of the above equations over 
any leaf also vanishes. \\

Substituting for $K$ in the above equations and {\it{specializing to
spherical topology}} (genus = 0) gives,

\begin{eqnarray}
2 \pi ~ & = & ~ \int_{\Sigma_2} (\pi\bar{\pi} + \frac{\underline{\delta}
\pi + \underline{\bar{\delta}} \bar{\pi}}{2} - {\cal{E}} - D\mu) - 
\int_{\Sigma_2} (\epsilon + \bar{\epsilon}) \mu ~~~, \\
0 ~ & = & ~ \int_{\Sigma_2} ( \underline{\delta} \pi \pm 
\underline{\bar{\delta}} \bar{\pi} ) 
\end{eqnarray}

We also deduce that,

\begin{eqnarray}
-(K - \bar{K}) ~ = ~ \Psi_2 - \bar{\Psi_2} & = &
- \{ \delta \pi - (\bar{\alpha} - \beta)\pi \}
+ \{ \bar{\delta} \bar{\pi} - (\alpha - \bar{\beta)} \bar{\pi} \}
\nonumber \\
& = & ~ \underline{\delta}\pi - \underline{\bar{\delta}}\bar{\pi}
\end{eqnarray}

Note that all these equations are manifestly invariant under 
rotations. Also observe that if $\Psi_2$ is {\em {not}} real, then 
{\em{$\pi$ can not be set to zero}}. \\

Since $\Psi_2$ is invariant under boost, scaling and rotation transformation
when $\kappa = 0$, its value has a physical meaning. incidentally,
$\Lambda, \Phi_{11}$ are also invariant under these transformations.
Thus, the complex curvature $K$ of the leaves is an invariant. \\ 

Above we obtained imaginary part of the complex curvature in terms of
derivatives of $\pi, \bar{\pi}$ in the gauge chosen. Since $K$ is invariant 
under boost transformations, the right hand side of eqn. (14) above, 
must also be invariant under residual boost transformations. This implies 
that we must have,

\begin{equation}
\underline{D} \pi ~ \equiv ~ D \pi + (\epsilon - \bar{\epsilon}) \pi = 0
\end{equation}

Now the equations ($b + \bar{d}$) of item 5 of the Appendix A imply that, 

\begin{equation}
\bar{\delta} (\epsilon + \bar{\epsilon})  = D\pi + ( \epsilon - \bar{\epsilon} )\pi = 0 
\end{equation}

Thus, $\epsilon + \bar{\epsilon}$ is constant on each leaf but it may
vary from leaf to leaf. This now limits the scaling freedom to scaling
by a factor which is constant on each leaf. We may now exhaust this freedom
by setting $\epsilon + \bar{\epsilon}$ to a constant on $\Delta$. Alternatively
we may note that invariance of $\underline{D}\pi = 0$ condition under 
residual boost transformations implies that,

\begin{equation}
D (\epsilon + \bar{\epsilon}) = 0
\end{equation}

We have therefore already got that real part of $\epsilon$ is a constant
on $\Delta$. This of course reduces the scaling to scaling by
a constant factor. \\

Thus at this stage we have $\kappa, \rho, \sigma, (\alpha + \bar{\beta}
- \pi), \Phi_{00}, \Phi_{01}$ are zero and $\mu$ is real. As a consistency
of the equations with the gauge choice, we also deduced that $\epsilon +
\bar{\epsilon}$ is constant over the horizon.  We have neither 
required $\nu$ to be zero nor that $\mu$ is strictly negative (positive). 
Physically we have already captured a {\it {geometrical}} property of $\Delta$ 
that it is {\it {potentially}} foliated by marginally trapped surfaces in a 
physical space-time. \\

We already have a definition of area namely the area of a leaf with
respect to the induced metric on $\Sigma_2$ and that this area is
``constant". What could be candidate for a ``surface gravity"? In the
usual case of stationary black holes, it is the acceleration at the
horizon of the killing vector normal to the horizon, with suitable 
normalization of the stationary killing vector at infinity.
Presently we don't have any stationary killing vector. The topology of
$\Delta$ however suggests that $\ell$ serves to define an evolution
along $\Delta$. Thus, its acceleration, $\tilde{\kappa} \equiv (\epsilon
+ \bar{\epsilon})$ is a natural candidate. Indeed, as we have seen above,
{\it{ $\tilde{\kappa}$ is constant over $\Delta$}}! Identifying
$\tilde{\kappa}$ with surface gravity (modulo a constant scale factor to
be fixed later) we already {\it{ have the zeroth law}}. \\

The rest of the logic is similar to \cite{isolated}. The Bianchi
identity implies, 

\begin{equation}
(D + \tilde{\kappa})(D \mu) = 0
\end{equation}

{\underline{Remarks:}}. 
\begin{enumerate}
\item We have deduced the zeroth law using just the two conditions
(non-expanding horizon), use of foliation and use of residual local
Lorentz transformation. The notion of weak isolation \cite{distorted} 
is not necessary. 
This is an alternative mentioned by Ashtekar et
al in \cite{distorted}. Although we have fixed the residual scaling 
freedom to a constant scaling only, we are still left with full rotation 
freedom and residual boost freedom. \\

\item All of the above works with any initial choice of $\ell$. In the 
process though we got the scaling to be restricted to a constant scaling 
only. Thus if we regard two $\ell$'s as equivalent if they differ by a 
constant non-zero factor, then all of the above holds for any given 
equivalence class of $\ell$. We could however begin with an equivalence 
class such that $D\mu = 0$. As long as an initial $\mu$ is non zero we can 
always do a local scaling transformation to the new $\mu$ to satisfy 
$D^{\prime}\mu^{\prime} = 0$. This will fix the initial $\ell$ and hence 
its equivalence class. Thus it is possible to choose a unique equivalence 
class of $\ell$ such that $D\mu = 0$. Note that this condition is preserved 
by the residual transformations. At this stage we do not need to 
make such a choice though we will use this towards the end of this
section. \\

\item We have obtained two constant quantities, surface gravity and area, 
associated with $\Delta$.  We could obtain $\tilde{\kappa}$ in terms of 
integrals of certain expressions over a leaf via eqn. (12). Notice that in 
the absence of $\mu$ being constant on a leaf, Gauss-Bonnet does not give 
$\tilde{\kappa}$ directly in terms of the area. \\

\end{enumerate}

\subsection{Freely Specifiable Spin Coefficients}

As noted earlier, the zeroth law refers to a single solution while the
first law refers to the class of solutions. To gain an understanding of
such a class, we now proceed to identify freely specifiable spin
coefficients corresponding to (non expanding) horizons. For this
of course we have to choose a suitable gauge. \\

Since we view $\Delta$ as a sub-manifold of a solution, we now consider
excursion off-$\Delta$. We still continue with an arbitrary initial
choice of the null normal vector field and initial $\Sigma_2$ (now
restricting to spherical topology). On $\Delta$ we have constructed null
tetrads and therefore have $n^{\mu}\partial_{\mu}$ defined. Consider
geodesics specified by points on $\Delta$ and the $n^{\mu}\partial_{\mu}$. 
We need only infinitesimal geodesics to go infinitesimally off-$\Delta$.
In this neighbourhood we construct null tetrads by parallel transporting
the null tetrads from $\Delta$, $n\cdot\nabla$(tetrad)$ = 0$. The
equations of item 2 from appendix A then immediately gives, $\gamma,
\nu, \tau$ to be zero off-$\Delta$ and therefore, by continuity also on
$\Delta$. \\

Thus, on $\Delta$, we have six of the twelve spin coefficients to be
zero, namely, $\gamma, \kappa, \nu, \rho, \tau, \sigma$. Furthermore we
have real part of $\epsilon$ to be a constant, $\mu$ to be real and 
$\pi = \alpha + \bar{\beta}$. \\

Now one can always choose a gauge such that imaginary part of $\epsilon$ is
zero on $\Delta$ (see the transformation equations of appendix A). This
has two consequences. Firstly it reduces the rotation freedom, so far
unrestricted, to rotations by parameter which is constant along the null
generators. It is still local along the leaves. Secondly, the eliminant
equation of appendix implies that $\alpha - \bar{\beta}$ is constant
along the null generators of $\Delta$. Thus both the combinations,
$\alpha \pm \bar{\beta}$ are now constant along the null generators and
$\epsilon$ is a real constant. Now only $\alpha \pm \bar{\beta}, \mu$ and
$\lambda$ are non trivial functions on $\Delta$. Are all of these freely
specifiable?\\

Not yet! The spin coefficients have still to satisfy the Einstein
equations. Only the Ricci scalars enter in these equations and Weyl
scalars can be thought of as derived quantities via the 18 complex equations 
of \cite{chandra}. The eliminant equation directly give relations among the
spin coefficients. \\

In the Appendix A, we have collected equations from \cite{chandra}.
These equations use the conditions derived from the non expanding
horizon conditions discussed above. Since we are interested in quantities 
defined on $\Delta$, the equations involving $D'$ derivatives are omitted. 
These can be understood as specifying the derivatives off $\Delta$ in terms 
of quantities specified on $\Delta$. \\

We see immediately that the eliminant equations give no conditions on
the non trivial spin coefficients mentioned above. From the full set of
the 18 equations of \cite{chandra} one can see the following. All quantities 
are evaluated on $\Delta$.  \\

\begin{itemize}
\item $\Phi_{00}$ appears in equation (a) which is identically true. 
\item $\Phi_{01}$ appears in equations (c, d, e, k) together with
$\Psi_1$. These serve to give $\Psi_1 = 0$ and $D^{\prime} \kappa = 0$.
This in turn implies that $\kappa$ is zero in an infinitesimal
neighbourhood of $\Delta$. 
\item $\Phi_{02}$ appears in equations (g, p). These serve to give
$D^{\prime} \sigma = - \Phi_{02}$ and also 
\begin{equation}
\underline{D} \lambda - \underline{\bar{\delta}} \pi - \pi^2 +
\tilde{\kappa} \lambda = \Phi_{02}
\end{equation}
This is a non trivial condition among the non vanishing coefficients and
can be thought of as a differential equation for $\lambda$ given the
other quantities.
\item $\Phi_{12}$ appears in equations (i, m, o, r) together with
$\bar{\Psi}_3$. Use (r) to eliminate $\Psi_3$. The remaining equations
show that $D^{\prime} (\pi - \alpha - \bar{\beta}) = 0$ and $\Phi_{12}$
determines only the off-$\Delta$ derivatives. One consequence of these
is that $\pi = \alpha + \bar{\beta}$ in a {\it{neighbourhood}} of $\Delta$.
\item $\Phi_{22}$ appears only in equation (n) and specifies $D^{\prime}
\mu$. The equation being real, it follows that $\mu$ is {\it{real}} in
a neighbourhood.
\item $\Phi_{11}, \Lambda$ appear in combinations with $\Psi_2$
in equations (f, h, l, q). One can use (l) to determine $\Psi_2$ and
eliminate the complex curvature of leaves, $K$, from the remaining
equations. Two of the equations then determine $D^{\prime} \epsilon$
and $D^{\prime} \rho$ while the remaining one gives,
\begin{equation}
D \mu - \underline{\delta} \pi -\pi \bar{\pi} + \tilde{\kappa} \mu =  
\alpha \bar{\alpha} + \beta \bar{\beta} - 2 \alpha \beta - \delta \alpha
+ \bar{\delta} \beta + \Phi_{11} + 3 \Lambda
\end{equation}
Again this can be thought of as a differential equation for $\mu$ in
terms of the remaining quantities. Notice that the equation (q) implies
that if $\Psi_2$ is real then, $\rho$ is real in the neighbourhood. 
\item Lastly equation (j) just gives $\Psi_4$ in terms of $D^{\prime}
\lambda$. 
\end{itemize}

In effect we have obtained two (differential) conditions on two of the
non trivial coefficients, $\mu$ and $\lambda$ and are left with just
$\alpha \pm \bar{\beta}$ as freely specifiable. Since these are constant
along the null generators, these need to be given only on a leaf. \\

The differential equations for $\lambda, \mu$ can be `reduced' further.
Recall the remark about the initial choice of the null normal vector
field. Generically we can choose it to be such that $D \mu = 0$. This is
consistent with the differential equation due to Bianchi identity and
the zeroth law. This fixes a unique equivalence class of null normals.
We could consider imposing $\underline{D}\lambda = 0$ condition also. 
If $\Phi_{02}$ is zero, as is the case for vacuum, Einstein-Maxwell and
Einstein-YM systems, the condition is consistent with the differential
equation. If it is non zero, then consistency with the differential
equation requires $D\Phi_{02} = 0$ which in turn via Bianchi identity (e)
requires $D^{\prime} \Psi_0 = 0$. Constancy of $\mu$ and $\lambda$ along
the generators is precisely the extra condition for isolated (as opposed 
to weakly isolated) horizons that is imposed by Ashtekar et al
\cite{distorted}. \\

Without the constancy of $\mu, \lambda$ along the generators, we
see that $\pi = (\alpha + \bar{\beta}), (\alpha - \bar{\beta}), \mu$ and
$\lambda$ may be freely specified on a leaf. $\tilde{\kappa}$ may then
be determined via the Gauss-Bonnet integral. Not all these specifications
give different (non-expanding) horizons since we still have good deal of
residual transformations apart from the initial choices of $\ell$ and
foliation. \\

With the constancy of $\mu, \lambda$ along generators imposed, we fix
the initial equivalence class of $\ell$, determine $\lambda$ and
$\tilde{\kappa} \mu$ in terms of $\alpha \pm \bar{\beta}$ and $\Phi_{02}$
specified on a leaf. The residual rotation can be fixed completely by making
$\alpha - \bar{\beta}$ real. We are then left with only the constant scaling 
and the dependence on the initial foliation (residual boosts). Notice that 
when $\epsilon$ is real, $\alpha - \bar{\beta}$ is invariant under boost 
transformations. Thus our identification of freely specifiable spin
coefficients is `gauge invariant'. The residual transformations now just
serve to demarcate equivalent {\it{isolated}} horizons. \\

\underline{Remarks} 
\begin{enumerate}
\item Apart from seeing the role of Einstein equations via
the 18 equations of \cite{chandra}, we also determined the Weyl scalars.
Of the five Weyl scalars, only $\Psi_2, \Psi_3$ are determined by freely
specifiable coefficients on a leaf and Ricci scalars. Thus the (real)
curvature of the leaves, and hence the metric on the leaves is also 
determined. The freely specifiable data consists of just one complex
function, $\pi$, and one real function, $\alpha - \bar{\beta}$, on a leaf.\\

\item It is apparent from our analysis and also proved by Lewandowski 
\cite{lewandowski} that the space of solutions admitting isolated horizons 
is infinite dimensional. In the above  analysis we have addressed the 
freedom of local Lorentz transformations naturally present in a tetrad 
formulation. We also have the diffeomorphism invariance though. Thus two 
sets of data related by a diffeomorphism on a leaf must be regarded as giving 
the `same' isolated horizon. This is of course implicit in an analysis 
based on an `initial value problem' formulation \cite{lewandowski}. The 
corresponding diffeomorphism classes must be characterized by diffeomorphism 
invariants. Integrals of scalars (eg invariant combinations of spin 
coefficients) over the leaf are just one such set of invariants. We will
obtain a few of these in the next section. \\
\end{enumerate}

To summarize: We have shown that every solution of Einstein-matter
equations with matter satisfying the dominant energy condition and admitting 
non-expanding horizon admits foliations and a corresponding choice of null 
tetrads modulo constant scalings such that the free data consists of one 
complex function $\pi$ and one real function $\alpha - \bar{\beta}$. 
Furthermore with $\tilde{\kappa}$ being identified as (unnormalized) surface 
gravity, the zeroth law holds for all such solutions.\\

In the next section we will consider the symmetry properties of isolated
horizons and see that if leaves admit at least one isometry, there is a
unique choice of foliation and the equivalence is reduced to that
implied by constant scaling only.

\section{Constants Associated with Isolated Horizons}

In this section we discuss symmetries of isolated horizons and
corresponding ``conserved" quantities. This discussion is carried out
in terms of invariant quantities so that the conclusions are not tied to
any particular gauge choice. Under certain condition we also see how this
helps to to fix a unique foliation. \\

\subsection{Symmetries of $\Delta$}

In the previous section  we naturally obtained two constant quantities 
associated with $\Delta$, the surface gravity and the area. Noting that 
quantities constructed out of the spin coefficients and their
derivatives {\it{and}} which are invariant under the residual transformations
are {\it{physical}} characteristics of a {\it {given}} isolated horizon, 
$\Delta$. An example is the complex curvature $K$ defined above. To look for 
further characteristic quantities associated with a given $\Delta$ we analyze 
the symmetries of these horizons.\\

Consider a fixed solution containing $\Delta$. This is characterized by a 
set of physical quantities. By definition, a symmetry of $\Delta$ is a 
diffeomorphism of $\Delta$ which leaves these quantities invariant. Since the 
induced metric on leaves is one such quantity, a symmetry must be an {\it{
isometry of the leaves}}. However, there are further invariant quantities 
built from ingredients other than the induced metric on leaves, eg $(K - 
\bar{K})$. We now look for further such invariant quantities. \\

Since residual transformations permit change of a foliation, it is desirable 
to make use of quantities invariant under these. Thus in order to classify 
the symmetry classes and distinguish a rotating case, we look for invariant 
forms and vector fields on $\Delta$. Any vector field (or form) can be
expressed as an expansion in terms of the appropriate tetrad basis. Depending 
on the transformation properties of the expansion coefficients under residual 
Lorentz transformations, these expansions will (or will not) preserve their 
form. By invariant vectors (forms) we mean form invariance under residual 
transformations. The general coordinate transformations of course play 
{\it{no}} role in the discussion. The use of such `invariant' quantities
frees us from having to keep track of the particular choices of tetrad
bases. The demand of form invariance puts restrictions on the expansion 
coefficients which can then be taken in a convenient manner as seen below. \\

Noting that the tangent space of $\Delta$ is spanned by the {\it {tangent}}
vectors $\ell, m, \bar{m}$ while the cotangent space is spanned the
{\it {cotangent}} vectors $n, m, \bar{m}$, it is easy to see that a
generic invariant vector field $X$ on $\Delta$ is parameterized as:

\begin{equation}
X ~ \equiv ~ \frac{1}{- \tilde{\kappa}}( \zeta + \pi \bar{C} + \bar{\pi} C
)\ell + C m + \bar{C} \bar{m} \\
\end{equation}

where, $C$ transforms as $\pi$ under rotation, is invariant
under boost while $\zeta$ is real and invariant under both sets of
transformations. These are our candidates for generating symmetries of
$\Delta$. \\

The condition that $X$ be a Killing vector (of the metric on $\Delta$
and {\it{not}} the four dimensional space-time metric) requires, in terms 
of the rotation covariant derivatives (see appendix A):

\begin{equation}
\underline{\delta} \bar{C} ~ = ~ 0 ~,~ \underline{\bar{\delta}} C ~ = ~
0 ~,~ \underline{\delta} C + \underline{\bar{\delta}} \bar{C} ~ = ~ 0.
\end{equation}

The third  of the above equations can be solved identically by setting
$C \equiv i \bar{\delta} f$ where $f$ is an invariant function while the
first two require $f$ to satisfy $\underline{\delta}^2 f = 0 = \underline{\bar{\delta}}^2 f$. \\

While looking for invariant 1-form, one should note that restricted to
forms , the boost transformations have the $\ell$ dependent terms
dropped. Then there are two types of such 1-forms (real and space-like):

\begin{eqnarray}
\omega & \equiv & -\tilde{\kappa} n + \pi m + \bar{\pi} \bar{m} ~,~  ~~~
\omega(X) ~ = ~ \zeta ; \\
\tilde{\omega} & \equiv & \tilde{C} m + \bar{\tilde{C}} \bar{m} ~,~ 
~~~~~~~~~~~~ \tilde{\omega}(X) ~
= ~ - ( C \bar{\tilde{C}}  + \bar{C} \tilde{C} )\nonumber
\end{eqnarray}


Since the 1-form $\omega$ is invariant and built out of spin
coefficients, it is a physical characteristic of $\Delta$ and thus a
symmetry generating invariant vector field must leave this invariant,
${\cal{L}}_{X} \omega = 0$. \\

This immediately implies that $D(\zeta) = 0$. It also gives two differential 
equations for $\zeta$, namely,

\begin{equation}
\delta \zeta = -\bar{C} (K - \bar{K}) ~~~~~,~~~~~
\bar{\delta} \zeta = C (K - \bar{K}) 
\end{equation}

This has several implications. Firstly, $X(\zeta) = 0$ i.e. $\zeta$ is
constant along the integral curves of $X$. Secondly, taking $D$ of the
equations, using the commutators of the derivatives and Bianchi identity 
implies that ${\underline{D}}C = 0$. Finally, The integrability conditions for 
these equations, which require that $X(K - \bar{K}) = 0$ and $\underline{
\delta}C + \underline{\bar{\delta}}\bar{C} = 0$, are automatically true by our 
definition of symmetry. It follows then that the solution for $\zeta$ is 
unique up to an additive constant. Note that when $K$ is real, $\zeta$ must 
be a constant. These integrability conditions will also allow us to set up 
an adapted ($\theta, \phi$) coordinate system on leaves. \\ 

One can construct an invariant vector field $Y$ (not necessarily a
symmetry generator) which is orthogonal to a symmetry generator $X$ by taking 
$C$ going to $i \Phi C$ where $\Phi$ is a real and invariant function. We 
would also like to use a symmetry generator along $\ell$. Thus we define two 
further invariant vector fields, 

\begin{eqnarray}
Y & \equiv & \frac{1}{- \tilde{\kappa}}( \zeta - i \Phi ( \pi \bar{C} - \bar{\pi} C)
)\ell + i \Phi (C m - \bar{C} \bar{m} ) \\
Z & \equiv & \frac{1}{- \tilde{\kappa}} \ell
\end{eqnarray}

Evidently, $X, Y$ are real, space-like and mutually orthogonal (for all 
$\Phi$). $Z$ is of course a null symmetry generator and is orthogonal to $X$ 
and $Y$. Observe that $\omega(X) = \zeta = \omega(Y), \omega(Z) = 1 $. \\ 

We would now like to have all these vector fields to commute so that 
parameters of their integral curves can be taken as coordinates for $\Delta$.
While commutativity of $X, Z$ is a statement about the nature of the
isometry algebra, commutators involving $Y$ are just stipulations on $Y$
which in no way affect properties of $X$ or $Z$. \\

The commutativity of the vector fields $X, Y$ with $Z$ requires:

\begin{equation}
\underline{D} \zeta ~ = ~ 0  ~~~ , ~~~  \underline{D} C ~ = ~ 0  ~~~ , ~~~ 
\underline{D} \Phi ~ = ~ 0   
\end{equation}

Note that commutativity of $Z, X$ is already implied by the eqn. (24)
while that of $Z, Y$ is a condition on $\Phi$. These enable us to construct 
the invariant vector fields as follows. For any leaf, assume we could find a 
real function $f$, satisfying $\underline{\delta}^2 f = 0$. Extend it to 
$\Delta$ by Lie dragging by $\ell$. This implies that $f$ so constructed, 
satisfies the double derivative condition on all the leaves. Furthermore, 
these conditions themselves are invariant under residual transformations. We 
could similarly define $\Phi, \zeta$ on $\Delta$. \\

The commutativity of $X, Y$ with $X$ a Killing vector (of leaves) requires 
$X(\Phi) = 0$ and

\begin{equation}
X(\zeta) -  Y(\zeta) ~ = ~ 2 i C \bar{C} \Phi (K - \bar{K}) \\
\end{equation}

The above equation automatically holds due to eqn (24). The only
additional information we have obtained is that the as yet arbitrary $\Phi$ 
function is constant along integral curves of $X$. There are no conditions 
implied on $X$ due to the demands of commutativity.\\

Several consequences can be derived now. \\

The integral curves of $X$ - the Killing orbits - in general leave a
leaf. We could however {\em {choose}} a foliation such that these orbits
are confined to leaves. This means that the coefficient of $\ell$ must
be made zero. This can be effected by a boost transformation. This
still leaves a one parameter freedom of boost residual transformation.
This can be fixed by demanding that integral curves of $Y$ be similarly
confined to leaves. The boost parameter effecting this is given by,

\begin{equation}
a + \frac{\bar{\pi}}{\tilde{\kappa}} = \frac{i \zeta}{2 \tilde{\kappa}
\Phi C} ( 1 + i \Phi )
\end{equation}

That the parameter $a$ satisfies the conditions of being residual
transformation requires $X(\zeta) = 0$ which is already seen to be true. 
For the transformation parameter to be well defined, it is necessary
that $\zeta$ vanishes where ever $C\Phi$ does. Since $C$ must vanish at
at least one point on a leaf, $\zeta$ must also vanish at at least one
point. Thus, when $K$ is real which implies that $\zeta$ is a constant,
$\zeta$ in fact must be zero.\\

Note that this fixes the boost freedom completely. The left hand side is 
nothing but the transformed value of $\bar{\pi}$ divided by the surface 
gravity. Thus in effect, this boost transformation has fixed for us a 
particular foliation. In this foliation we have the $X, Y$ vector 
fields purely tangential to the leaves. Consequently, we call the vector
field $X$ as a ``rotational" symmetry generator. \\

In this foliation $\zeta = -( \pi \bar{C} + \bar{\pi} C )$ and we see
that {\it {$\zeta$ is zero iff there exist a gauge in which $\pi$ is zero.}} 
This fact will be used in defining the angular momentum in terms of $\zeta$. \\

Since the leaves are compact, the Killing vector field is complete. Suppose
for the moment that its orbits are closed curves. Provided that the ranges 
of the Killing parameter is the same for all the orbits, we can adjust this
range to be $2 \pi$ by a constant scaling of $X$ and identify the
Killing parameter to an angular coordinate $\phi$. This will genuinely
make the spatial Killing vector correspond to axisymmetry. When can
this be done? \\

First we need to argue that orbits of the Killing vector, $X$, are
closed. Every vector field on $S^2$ will have at least one zero (or
fixed points). These fixed points can be classified in elliptic, hyperbolic 
etc by standard linearizations \cite{fixedpoints}. The third of the eqn. (22) 
implies that $X$ is `area preserving' and thus has zero divergence. This 
implies that its zeros are all elliptic and hence orbits in the vicinity are 
closed. Since the orbits can not intersect, all orbits in facts must be closed. 
The vector field $Y$ by contrast is {\it{not}} divergence free and thus its 
orbits are {\it{not}} closed. \\

For an integral curve $\gamma$ of $X$, the line integral of the
invariant 1-form $\omega$ is equal to the integral of $\omega(X) = \zeta
$ along $\gamma$. Since $\zeta$ is constant along such curves, its
integral will be $\zeta(\gamma) \times I(\gamma)$, where $I(\gamma)$ is the
range of the Killing parameter along $\gamma$. Now one can consider a
family of such $\gamma$'s labeled by an infinitesimal integral curve
of $Y$. Differentiate the integrals w.r.t. the parameter along $Y$. On
the left hand side use Stokes theorem together with $d\omega \ = \ (K -
\bar{K}) m \wedge \bar{m} $ while on the right hand side use Taylor expansion. 
This gives the left hand side as,

\begin{equation}
\oint_{\gamma(\beta + \delta \beta)} \omega - \oint_{\gamma(\beta)} \omega 
= \int_{cyl} d \omega = \int_{cyl} (K - \bar{K}) m \wedge
\bar{m} \sim - 2 i \Phi C \bar{C} (K - \bar{K})(\beta) I(\beta) \delta
\beta.
\end{equation}

Here the surface integral is over an infinitesimal cylinder formed by
the family. In the last step of course we have used the mean value
theorem. The right hand side gives, 


\begin{equation}
\{ Y(\zeta) I(\beta) + \zeta(\beta) Y(I(\beta)) \} \delta\beta
\end{equation}
 
The equation (28) satisfied by $\zeta$ then implies that the parameter range
does not vary along $Y$ and hence is a constant. We can now introduce
the azimuthal angle $\phi$. Clearly $\zeta, \Phi$ are now independent of
$\phi$. \\

We have used the vector field $Y$ to show that azimuthal coordinate can
be introduced. This is independent of the choice of $\Phi$. Can we also
introduced the polar angle $\theta$? The answer is yes as seen below.
\\

Every vector field on $S^2$ must vanish at at least one point. $X$
vanishes precisely when $C = 0$. If $\Phi$ is non singular then $Y$ also
vanishes precisely where ever $X$ vanishes. One can ``decompactify" the
leaf by removing such a point and see that $C$ must vanish at one and
only one more point. Further the integral curves of $Y$ must ``begin" and
``end" at these two points. This may also be seen via the Poincare-Bendixon 
theorem \cite{fixedpoints}. Further, since $X, Y$ commute, the 
diffeomorphisms generated by $X$ take orbits of $Y$ to orbits of $Y$. 
Choosing a ``meridian" we could suitably adjust $\Phi$ and hence select
a $Y$ so that the parameter along this curve ranges from 0 to $\pi$. Lie 
dragging by $X$ then defines these for other longitudes. This way we can 
introduce the standard spherical polar angles on the leaf (boost gauge fixed).
The metric can also be written down as:

\begin{equation}
ds^2 = -2C \bar{C} ( \Phi^2 d\theta^2 + d\phi^2 )
\end{equation}

That such a choice of $\Phi$ is possible can be seen by construction.  For 
the above form of the metric, one can compute the Ricci scalar in terms of 
$\Phi$. We also have $(K + \bar{K})$ as the curvature of the leaves. 
Equating these two gives a first order differential equation for $\Phi$. 
Its smooth behaviour on the leaf excepting the poles, fixes the constant 
integration and determines $\Phi$ in terms of $C$ and $(K + \bar{K})$. For 
the Kerr-Newman family it reproduces the precise metric components. Thus 
existence of an invariant Killing vector (spatial), permits us to determine 
both $\zeta$ and $\Phi$ and also allows us to introduce an adapted set of 
coordinates $\theta, \phi$ on the spherical leaves. Note that while on each 
of the leaf we can introduce these coordinates, the choice of prime meridians 
on different leaves is arbitrary. These could be related by diffeomorphisms 
generated by $Z$. \\

\underline{Remark:} Commutativity of the invariant vector fields, $X, Y,
Z$, not only allows us to introduce spherical polar coordinates on
leaves, it also allows us to introduce coordinates on $\Delta$ itself.
With the excursion off-$\Delta$ defined via parallel transport along
geodesics defined by $n^{\mu}\partial_{\mu}$, one can naturally
introduce coordinates in the neighbourhood also. \\

On a spherical leaf, we can have either none, one or three isometries
i.e. either 1) {\it {no}} such $f$, or 2) precisely {\it{ one}} such $f$ , 
or 3) precisely {\it{three}} such $f$'s. If we found two such functions 
(two Killing vectors), then their commutator is either zero or is also a
Killing vector. On $S^2$ we can't have two commuting Killing vectors,
therefore the commutator must be non zero and a Killing vector. One can
consider the commutator $X^{\prime\prime} \equiv \left[ X, X^{\prime}
\right] $ of two Killing vectors and see that it is also an invariant Killing 
vector provided the $\zeta^{\prime\prime}$ satisfies the same conditions. 
This is possible only if $K - \bar{K} = 0$. Thus, as expected, we see that 
maximal symmetry is possible only if $\Psi_2$ is real. In this case of
course one does not expect any rotation. The converse need not 
be true as shown by the ``distorted" horizons \cite{distorted}. \\

\underline{Remark:} We have fixed the boost freedom by demanding $X, Y$
be tangential to a leaf. When we have three such pairs of vector fields
are all these automatically tangential to the same leaf? The answer is
yes. Since three isometries implied $K$ is real, all the $\zeta,
\zeta^{\prime}, \zeta^{\prime\prime}$ are constants and hence zero.
Furthermore tangentiality of any one implies $\pi = 0$. Therefore all
the three Killing vectors are tangential to the same leaf.\\

In the case of single isometry, we have both the possibilities namely 
$K - \bar{K}$ is zero or non zero. These two can be seen intuitively as
hinting that while a ``rotating" body is expected to be distorted (bulged) 
a distorted body need not be rotating. Since maximal symmetry implies `no
rotation' and also that $K$ is real, we label the two cases as:\\

{\it {$\Delta$ is non-rotating iff $K$ is real (i.e. $\Psi_2$ is real) and is
rotating iff $K$ has non zero imaginary part.}} \\

We have already noted that in non-rotating case $\zeta$ must be zero. 
This means that $\pi$ can be transformed to zero. One could also see this 
directly from the equations of residual transformations that it is possible 
to choose a gauge such that $\pi = 0$. For the rotating case of course $\zeta$ 
is non zero and $\pi$ is also non zero in every gauge. \\

\subsection{``Conserved" quantities}

While $\zeta$ introduced above, could be non zero, it is some function of 
the adapted coordinate $\theta$. Its integral over the leaf is clearly a 
constant which is zero iff $K$ is real. This integral is therefore a candidate
for defining angular momentum. Indeed, the explicit example of Kerr space-time, 
discussed in appendix-B, shows that the candidate agrees with the angular 
momentum of the Kerr space-time. It also provides a proportionality factor. 
Unlike the Kerr black hole however, this angular momentum is defined in terms 
of quantities intrinsic to $\Delta $. For Kerr-Newman solution also, one can 
find $\zeta$ and its integral. Interestingly, one does {\em{not}} get the usual,
total angular momentum ($J = Ma$) of the Kerr-Newman space-time, but one gets
the total angular momentum {\em minus} the contribution of the electromagnetic
field in the exterior. Indeed, one can explicitly check that the integral of 
$\zeta$ is precisely equal to the Komar integral \cite{wald} evaluated at the 
event horizon. This justifies the case for taking the following as the angular
momentum of an isolated horizon.\\

Thus we define, 

\begin{equation}
J \equiv - \frac{1}{8\pi}\int_{S^2} \zeta ( i m \wedge \bar{m} )
\end{equation}

Apart from the ``rotational" symmetry generator defined above, we see
that $gZ$ is also a symmetry generator provided $g$ is a constant. Following
identical logic as for the rotational symmetry generator, we see that
analogue of $J$, is proportional to the area. Thus area is the
`conserved' charge associated with $Z$. Unlike $X$ whose normalization
is fixed due to compactness of its orbits, $S^1$, normalization of $Z$
is not fixed.  Clearly arbitrary constant linear combinations of $Z$ and 
the well defined ``rotational" $X$ is a symmetry generator. Let us denote 
this as,

\begin{equation}
\xi_{a,b} \equiv ( - a Z - b X); ~~~ \tilde{\zeta}_{a,b} \equiv 
\omega(\xi_{a,b}) = -(a + b \zeta)  
\end{equation}

We have introduced the suffixes $a,b$ to remind ourselves that the
suffixed quantities depend on the arbitrary constants $a,b$. \\

We define, by analogy with $J$, a corresponding conserved quantity as,

\begin{equation}
\tilde{M}_{a,b} ~~ \equiv ~~ -\frac{1}{4\pi} \int_{S^2} \tilde{\zeta}_{a,b} 
~~ = ~~ \frac{a}{4\pi} \mbox{Area} + 2 b J 
\end{equation}

Note that unlike $\zeta$ which is determined via a differential equation
involving ($K - \bar{K}$), $\zeta_{a,b}$ is {\it {not}} completely
determined by ($K - \bar{K}$). A Smarr-like relation still follows because
of the symmetries we have and our parameterization of the general
isometry, $\xi_{a,b}$ in terms an arbitrarily introduced constants
$a, b$.\\

Interestingly, the usual Smarr relation for the Kerr-Newman family, with
$M, J$ replaced by the Komar integrals evaluated at the event horizon
has exactly the same form as above with $a = \tilde{\kappa}_{KN}$ and $b =
\Omega_{KN}$! The Maxwell contribution is subsumed in the Komar integrals at 
the horizons and there is no explicit `$Q \Phi$' term. \\

It is thus suggestive that $M_{a,b}$ be identifiable as the mass of the
isolated horizon. It is defined on the same footing as the angular momentum.
For $b \ne 0 $ the $\xi_{a,b}$ is a space-like isometry (analogue of stationary
Killing vector) as in the case of the Kerr-Newman family. However $a,b$ 
are {\it{arbitrary}} parameters. Further, while $b$ is the analogue of the
angular velocity, it is not clear why it must be {\it{non-zero}} when
and only when $J$ is non-zero.  \\

The notion of angular velocity is tricky though because on the one hand 
`rotation' must be defined with respect to some observers (in the usual
case asymptotic observers) on the other hand it is signaled by
imaginary part of $\Psi_2$ which is a local property. At present we have
not been able to resolve this issue. \\

{\underline{Remarks:}} 
\begin{enumerate}
\item The above candidate definitions are not the same as
those given by Ashtekar et al \cite{distorted}. There need not be any
inconsistency since our definitions need not enter a first law in the
usual manner. Some functions of our quantities may become the mass and
angular momentum which will enter a first law. \\

\item Our candidate identifications are based on analogy and consistency 
with the Kerr-Newman family. The symmetry generators, $Z, X$,  are invariant
vector fields. In particular they are insensitive to the constant
scaling freedom un-fixed as yet. Provided, the constants $a, b$, are
also invariants, the general symmetry generator $\xi_{a,b}$ will also be
an invariant vector field. If $a, b$ are proposed to be the surface
gravity and the angular velocity based on analogy and dimensional
grounds, these will be subject to the constant scaling freedom 
and one will have to face the normalization issue as done in
\cite{isolated,distorted}. If on the other hand $a, b$ (equivalently
$\xi_{a,b}$) could be directly tied to the space-time under consideration, 
and are calculable then a first law variation may be directly obtained. In 
the absence of a stationary Killing vector there does not appear to be any  
natural choice of $\xi_{a,b}$.
\end{enumerate}

Without relating the constants $a, b$ to any other intrinsic properties of 
$\Delta$ and/or normalizing them suitably, $M_{a,b}$ can not fully be 
identified as the mass of $\Delta$. Since we do not address the first law in 
this work, we do not pursue the identification further.\\

To summarize: In this section we assumed the existence of at least one
spatial symmetry of $\Delta$ and deduced candidate definition of angular
momentum and possible identification of mass. We also saw how this enables 
us to fix a unique foliation together with an adapted set of spherical polar 
angular coordinates. At present, we have not been able to obtain
a candidate definition of angular velocity. Whether at least one spatial
isometry must exist or not is also not addressed. These will be addressed
else where.

\subsection{The Special Case of Spherical Symmetry}

If a solution admitting isolated horizon has maximal symmetry for the
leaves, we have already seen that complex curvature must in fact be
real. Furthermore it must be constant due to maximal symmetry. Our gauge
choice fixing the foliation then has $\pi = 0$. Equations (19) and (8)
which determine $\lambda$ and $\mu$ then become,

\begin{equation}
\tilde{\kappa} \lambda ~ = ~  \Phi_{20} ~~~~,~~~~ \tilde{\kappa}\mu 
~ = ~ -K - {\cal{E}} ~ \Rightarrow ~ \tilde{\kappa}\delta \mu ~ = ~ -
\delta {\cal{E}}
\end{equation}

If $\lambda = 0$ can be shown, then equation (m) of item 5 of appendix A 
will give $\Psi_3 = \Phi_{21}$. Like wise $\delta{\cal{E}} = 0$ will give $\mu$ 
to be a constant. It appears that maximal symmetry alone does {\it{not}} imply
either $\lambda$ or $\delta \mu$ to be equal to zero. For the Maxwell and YM
matter, the form of the stress tensor together with energy condition
implies $\Phi_{02} = 0$ and then $\lambda = 0$ does follow. Since the
space-time itself is not guaranteed to be spherically symmetric in the
neighbourhood, while $\delta{\cal{E}} = 0$ is a reasonable condition, it
does not seem to be forced upon. If these two conditions however, are imposed, 
then the previous results on non-rotating and spherically symmetric
isolated horizons \cite{isolated} are recovered. Incidentally, in this case 
positivity of $K$ (spherical topology) and of ${\cal{E}}$ implies that
$\tilde{\kappa} \mu$ is negative. Thus for positive surface gravity,
$\mu$ must be negative implying marginal trapping.\\

\section{summary and conclusions}

In this work we have presented a general analysis of isolated horizons
by manipulating the basic equations of null tetrad formalism. By using
existence of foliations (causally well behaved solutions) and keeping
track of local Lorentz transformations we have shown that non-expanding
conditions (as opposed to weak isolation) are adequate to get the zeroth law. 
We also used a gauge
fixing procedure to identify freely specifiable spin coefficients. By
using invariant vector fields and 1-forms, we analyzed the symmetries of
isolated horizons and showed that if at least one spatial isometry
exists, then a unique foliation can be fixed. Further more a natural
choice of spherical polar coordinates exists for the leaves. We defined
associated `conserved' quantities which are consistent with the
Kerr-Newman family. In the appendix B we have illustrated our procedures
for the Kerr-Newman family. \\

We have not taken explicit examples of various matter sectors (except the
Kerr-Newman family). For the Einstein-Maxwell, Einstein-Yang-Mills with
or without dilaton has been discussed in \cite{isolated}. We have also
not invoked any action principle. Our hope was and still is to deduce
a first law {\it{without}} the use of an action principle. Action
formulation(s) and its use has already been discussed in detail in
\cite{isolated,distorted}. \\

The first law for isolated horizons is much more subtle. A very interesting
formulation and perspective is given by Ashtekar et al \cite{distorted}.
Here we will be content with some remarks. \\

The usual stationary black holes are parameterized by finitely many
parameters (3 for the Kerr-Newman family) and the first law also
involves only a few parameters. By contrast, the space of solutions
with isolated horizons is infinite dimensional and yet a first law is
expected to involve only a few parameters. Even granting the infinitely
many forms of the first law \cite{distorted}, each of these still
involves a few parameters only with area, angular momentum and charges
playing the role of independently variable quantities. Why only a few
quantities be expected {\it{a priori}} to enter a first law of mechanics
for isolated horizons is unclear. Ashtekar et al have incorporated 
asymptotic flatness in an action principle and used the covariant phase 
space formulation to deal with the space of solutions directly. They traced 
the existence of first law as a necessary and sufficient condition  for a 
Hamiltonian evolution on the covariant phase space of isolated horizons. 
It will be nicer to have a direct and quasi-local `explanation' of the first 
law. \\

{\underline{Acknowledgments:}} 

It is a pleasure to acknowledge the illuminating and fruitful 
discussions Abhay Ashtekar provided during his visit to Institute of
Mathematical Sciences in February, 2000. The stimulating discussions with
Madhavan Varadarajan, Richard Epp and Samuel Joseph of the Raman Research
Institute during a visit there are gratefully acknowledged. I wish to
thank Abhay Ashtekar for his critical comments on an earlier draft.

\newpage
{\underline{Appendix A}} {\bf {Summary of relevant equations for spin 
coefficients}} \\

In this appendix we collect together the relevant and useful formulae
from \cite{chandra}. The conventions are those of Chandrasekhar's
book. 

\begin{enumerate}
\item Null tetrad basis in the tangent and cotangent spaces: 

\begin{eqnarray*}
E_1 = \ell^{\mu}\partial_{\mu} ~\equiv D ~,~ E_2 = n^{\mu}\partial_{\mu}  
~\equiv D^{\prime} ~,~  E_3 = m^{\mu}\partial_{\mu} ~\equiv \delta
~,~  E_4 = \bar{m}^{\mu}\partial_{\mu} ~\equiv \bar{\delta} ~; \\
E^1 = n_{\mu}dx^{\mu} ~~~~~~,~  E^2 = \ell_{\mu}dx^{\mu}  
~~~~~~~~,~  E^3 = - \bar{m}_{\mu}dx^{\mu}  ~~~,~  E^4 = - m_{\mu}dx^{\mu} ~~~~~
\end{eqnarray*}

\item Covariant derivatives in terms of spin coefficients:

\begin{eqnarray*}
\ell_{\mu ; \nu} & = & (\gamma + \bar{\gamma}) \ell_{\mu} \ell_{\nu} +
(\epsilon + \bar{\epsilon})\ell_{\mu}n_{\nu} - (\alpha + \bar{\beta})
\ell_{\mu} m_{\nu} - (\bar{\alpha} + \beta) \ell_{\mu} \bar{m}_{\nu} \nonumber
\\
&   & -\bar{\tau}m_{\mu}\ell_{\nu} - \bar{\kappa}m_{\mu}n_{\nu} +
\bar{\sigma}m_{\mu}m_{\nu} + \bar{\rho}m_{\mu}\bar{m}_{\nu} \\
&   & -\tau \bar{m}_{\mu}\ell_{\nu} - \kappa \bar{m}_{\mu}n_{\nu} +
\rho \bar{m}_{\mu}m_{\nu} + \sigma \bar{m}_{\mu}\bar{m}_{\nu} \nonumber
\\
\nonumber \\
n_{\mu ; \nu} & = & -(\epsilon + \bar{\epsilon}) n_{\mu} n_{\nu} -
(\gamma + \bar{\gamma}) n_{\mu}\ell_{\nu} + (\alpha + \bar{\beta})
n_{\mu} m_{\nu} + (\bar{\alpha} + \beta) n_{\mu} \bar{m}_{\nu} \nonumber\\
&   & + \nu m_{\mu}\ell_{\nu} + \pi m_{\mu}n_{\nu} -
\lambda m_{\mu}m_{\nu} - \mu m_{\mu}\bar{m}_{\nu} \\
&   & +\bar{\nu} \bar{m}_{\mu}\ell_{\nu} + \bar{\pi} \bar{m}_{\mu}n_{\nu} -
\bar{\mu} \bar{m}_{\mu}m_{\nu} - \bar{\lambda} \bar{m}_{\mu}\bar{m}_{\nu} \nonumber \\
\nonumber \\
m_{\mu ; \nu} & = & \bar{\nu} \ell_{\mu} \ell_{\nu} +
\bar{\pi} \ell_{\mu}n_{\nu} - \bar{\mu} \ell_{\mu} m_{\nu} - \bar{\lambda}
\ell_{\mu} \bar{m}_{\nu} \nonumber\\
&   & - \tau n_{\mu}\ell_{\nu} - \kappa n_{\mu}n_{\nu} + \rho n_{\mu}m_{\nu} 
- \sigma n_{\mu}\bar{m}_{\nu} \\
&   & + ( \gamma - \bar{\gamma}) m_{\mu}\ell_{\nu} + (\epsilon - \bar{\epsilon})
 m_{\mu}n_{\nu} - (\alpha - \bar{\beta}) m_{\mu}m_{\nu} + 
(\bar{\alpha} - \beta) m_{\mu}\bar{m}_{\nu} \nonumber \\
\nonumber \\
\end{eqnarray*}

\item Local Lorentz transformations: 

These are given using $\kappa, \rho, \sigma, \Psi_0, \Psi_1, \Phi_{00}$
set equal to zero. These values are invariant under both the sets of
local Lorentz transformations. 

{\underline{Type-I (Boosts):}}

\begin{center}
\begin{tabular}{l}
$\ell' = \ell, ~~ n' = n + \bar{a} m + a \bar{m} + a \bar{a} \ell, ~~$ 
$m' = m + a \ell, ~~ \bar{m}' = \bar{m} + \bar{a} \ell $ ; \\
\\
$ \Psi_2' = \Psi_2 ~~,~~ \Psi_3' = \Psi_3 + 3\bar{a}\Psi_2 ~~,~~ 
\Psi_4' = \Psi_4 + 4\bar{a}\Psi_3 + 6 \bar{a}^2 \Psi_2 $; \\
\\
$\epsilon' = \epsilon ~~~,~~~ \tau' = \tau  $, \\ 
$ \pi' = \pi + 2 \bar{a}\epsilon + D\bar{a} ~~~,~~~ 
\alpha' = \alpha + \bar{a}\epsilon ~~~,~~~ \beta' = \beta + a \epsilon $, \\
$\gamma' = \gamma + a \alpha + \bar{a}(\beta + \tau) + a\bar{a}\epsilon
$,\\
$\lambda' = \lambda + \bar{a}(2\alpha + \pi) + 2\bar{a}^2\epsilon +
\bar{\delta}\bar{a} + \bar{a}D\bar{a} $, \\
$\mu' = \mu + a\pi + 2\bar{a}\beta + 2a\bar{a}\epsilon + \delta\bar{a} +
aD\bar{a} $, \\
$\nu' = \nu + a\lambda + \bar{a}(\mu + 2\gamma) + \bar{a}^2(\tau +
2\beta) + a\bar{a}(\pi + 2\alpha) + 2a\bar{a}^2\epsilon $\\
$ ~~~~~~ (D' + \bar{a}\delta + a\bar{\delta} + a\bar{a}D)\bar{a} $ ;
\end{tabular}
\end{center}

{\underline{Type-III (scaling and rotations):}} 

\begin{center}
\begin{tabular}{l}
$\ell' = A^{-1} \ell ~~,~~ n' = An ~~,~~ m' = e^{i\theta}m ~~,~~
\bar{m}' = e^{-i\theta}\bar{m} $, \\
\\
$\Psi_2' = \Psi_2 ~~,~~ \Psi_3' = A\Psi_3e^{-i\theta} ~~,~~ \Psi_4' =
A^2e^{-2i\theta}\Psi_4 $, \\
\\
$\mu' = A\mu ~~,~~ \tau' = e^{i\theta}\tau ~~,~~ \pi' = e^{-i\theta}\pi 
~~,~~ \lambda' = Ae^{-2i\theta}\lambda ~~,~~ \nu' = A^2e^{-i\theta}\nu $, \\
$(\epsilon + \bar{\epsilon})' = A^{-1} \{ (\epsilon + \bar{\epsilon})
-A^{-1}DA \} ~~~,~~~
(\epsilon - \bar{\epsilon})' = A^{-1} \{ (\epsilon - \bar{\epsilon})
 + iD\theta \} $, \\
$(\alpha + \bar{\beta})' = e^{-i\theta} \{ (\alpha + \bar{\beta})
-A^{-1}\bar{\delta}A \} ~~~,~~~
(\alpha - \bar{\beta})' = e^{-i\theta} \{ (\alpha - \bar{\beta})
 + i\bar{\delta}\theta \} $ 
\end{tabular}
\end{center}

\item Type-III rotation Covariant derivatives (``compacted covariant
derivatives"): 

These are defined for a quantity that transforms homogeneously under 
rotations. 

\begin{center}
\begin{tabular}{lcl}
$X^{\prime} = e^{in\theta} X $ & ~, ~ & n an integer \\
$ \underline{D}X ~ \equiv ~ DX - n (\epsilon - \bar{\epsilon}) X $ & ; & 
$ \underline{D}^{\prime}X^{\prime} ~ = ~ e^{in\theta} \underline{D}X $ \\
$ \underline{\delta}X ~ \equiv ~ \delta X + n (\bar{\alpha} - \beta) X $ & ; & 
$ \underline{\delta}^{\prime}X^{\prime} ~ = ~ e^{i(n + 1)\theta} \underline{\delta}X $ \\
$ \underline{\bar{\delta}}X ~ \equiv ~ \bar{\delta} X - n (\alpha - \bar{\beta}) X $ & ; & 
$ \underline{\bar{\delta}}^{\prime}X^{\prime} ~ = ~ e^{i(n - 1)\theta} \underline{\bar{\delta}}X $ \\
\end{tabular}
\end{center}

The corresponding commutators are given by:

\begin{center}
\begin{tabular}{lcl}
$\left[ \underline{\delta} , \underline{D} \right] X $ & = & 
$ ( \bar{\alpha} + \beta - \bar{\pi} ) \underline{D} X $ \\
$\left[ \underline{\bar{\delta}} , \underline{D} \right] X $ & = & 
$ ( \alpha + \bar{\beta} - \pi ) \underline{D} X $ \\
$\left[ \underline{\delta} , \underline{\bar{\delta}} \right] X $ & = & 
$ -n ( K + \bar{K} ) X $ 
\end{tabular}
\end{center}

\item Riemann tensor component equations: 

There are in all 18 such equations together with their complex
conjugates. But since we are interested in values only on the horizon,
we regard the equations involving $D'$ derivatives as specifying the
same in terms of values obtained on $\Delta$. These therefore give no
conditions on $\Delta$ itself. Three of these equations, equations (a),
(b) and (k) of \cite{chandra} have already been been implied by the
energy conditions, Raychoudhuri equation and the zero expansion
condition on the $\ell$ congruence. These give, $\Psi_0, \Psi_1,
\Phi_{00}, \Psi_{01}$ and $\Psi_{10} $ to be zero. This leaves us with 6 
equations, 2 of the eliminant equations and 2 from the Bianchi identities. 
These are listed below for convenience {\em{in terms of the ``compacted
covariant derivatives"}} where relevant. The equation labels refer to the 
equations from Chandrasekhar's book.

\begin{center}
\begin{tabular}{lcll}
$D\alpha - \bar{\delta}\epsilon $ & = & $ -\alpha (\epsilon - \bar{\epsilon})
+ \epsilon(\pi - \alpha - \bar{\beta}) $ & (eqn. d) \\
$D\beta - \delta\epsilon $ & = & $ \beta (\epsilon - \bar{\epsilon})
+ \epsilon(\bar{\pi} - \bar{\alpha} - \beta) $ & (eqn. e) \\
${\underline D}\lambda - \bar{\underline{\delta}}\pi $ & = & $ \pi^2 
- \lambda(\epsilon + \bar{\epsilon})  + \Phi_{20} $ & (eqn. g) \\
$D\mu - {\underline{\delta}}\pi $ & = & $ \pi \bar{\pi} 
- \mu(\epsilon + \bar{\epsilon})  + \Psi_2 + 2\Lambda $ & (eqn. h) \\
$\delta\alpha - \bar{\delta}\beta $ & = & $ \alpha\bar{\alpha} +
\beta\bar{\beta} - 2 \alpha\beta - \epsilon(\mu - \bar{\mu}) - \Psi_2 +
\Phi_{11} + \Lambda $ & (eqn. l) \\
${\underline{\delta}}\lambda - \bar{{\underline{\delta}}}\mu $ & = & $ 
\pi(\mu - \bar{\mu}) + \mu(\alpha
+ \bar{\beta}) - \lambda(\bar{\alpha} + \beta) - \Psi_3 + \Phi_{21} $ & 
(eqn. m) \\
\end{tabular}
\end{center}

\item Eliminant equations :

\begin{center}
\begin{tabular}{lcll}
$\delta(\alpha + \bar{\beta} - \pi) - 
\bar{\delta}(\bar{\alpha} + \beta - \bar{\pi}) $& = &$ 
- D(\mu - \bar{\mu}) - 2(\alpha\beta - \bar{\alpha}\bar{\beta}) $ & \\
& & $ -(\bar{\alpha} - \beta)\pi + (\alpha - \bar{\beta})\bar{\pi} $ 
& (eqn. b$^{\prime}$) \\ 
$D(\bar{\alpha} - \beta) + \delta(\epsilon - \bar{\epsilon}) 
$&=&$(2\bar{\alpha} - \bar{\pi})(\epsilon - \bar{\epsilon}) $ 
& (eqn. f$^{\prime}$) \\
\end{tabular}
\end{center}
\item Bianchi identities (terms which vanish are dropped): 

\begin{center}
\begin{tabular}{lclr}
$ -D\Psi_2 - D'\Phi_{00} - 2D\Lambda $ & = & 0 & $~~~~~~~~~~$(eqn. b$^{\prime\prime}$) \\
$ -D(\Phi_{11} + 3\Lambda) - D'\Phi_{00} $ & = & 0 & $~~~~~~~~~~$(eqn. i$^{\prime\prime}$) 
\end{tabular}
\end{center}

\item Change of the metric on $\Sigma_2$ under its diffeomorphism:

The metric on $\Sigma_2$ is $- (m\otimes\bar{m} + \bar{m}\otimes
m)_{\mu\nu}$. We are interested in the Lie derivative of this metric
with respect to a vector field on $\Sigma_2$ projected back on to
$\Sigma_2$. For a vector field of the form, $X = Cm + \bar{C}\bar{m}$, 
one gets,
%

\begin{eqnarray*}
{\cal{L}}_{X} (m \otimes \bar{m} + \bar{m} \otimes m) ~ & = & ~
2 {\underline{\bar{\delta}}}C (m \otimes m) + 
2 {\underline{\delta}}\bar{C} (\bar{m} \otimes \bar{m}) + 
({\underline{\delta}}C + {\underline{\bar{\delta}}}\bar{C}) (m \otimes
\bar{m} + \bar{m} \otimes m) \\
{\cal{L}}_{X} (m \otimes \bar{m} - \bar{m} \otimes m) ~ & = & ~
({\underline{\delta}}C + {\underline{\bar{\delta}}}\bar{C}) (m \otimes
\bar{m} - \bar{m} \otimes m)
\end{eqnarray*}

\end{enumerate}

\newpage
{\underline{Appendix B}} {\bf {Example of Kerr-Newman family}} \\

In this appendix we describe the Kerr-Newman family in terms of the
spin coefficients. While these are given by \cite{chandra}, a slight
modification is needed. Further, these are also used to illustrate our
gauge fixing procedure. \\

Metric: 

\begin{equation}
ds^2 = \frac{\eta^2 \Delta}{\Sigma^2} dt^2 - (\frac{\Sigma^2
sin^2\theta}{\eta^2})(d\phi - \omega dt)^2 - \frac{\eta^2}{\Delta} dr^2
- \eta^2 d\theta^2 , ~~~~~~~~ \mbox{where}
\end{equation}

\begin{center}
\begin{tabular}{lclclcl}
$\eta^2 $ & $\equiv $ & $ r^2 + a^2 cos^2\theta$ & $~~,~~$ & $\Sigma^2 $ & 
$\equiv $ & $ (r^2 + a^2)^2 - a^2 sin^2\theta \Delta $ \\
$\omega $ & $\equiv $ & $ \frac{a(2Mr - Q^2_{\star})}{\Sigma^2} $ & $~~,~~$ & 
$\Delta $ & $\equiv $ & $ r^2 + a^2 - 2Mr + Q^2_{\star} $ \\
$\xi $ & $\equiv $ & $ r + i a cos\theta $ & $~~,~~$ &
$\bar{\xi} $ & $\equiv $ & $ r - i a cos\theta $ 
\end{tabular}
\end{center}

The area, surface gravity and angular velocity of the event horizon are given 
by ($r_+$ is the radius of the event horizon),

\begin{eqnarray}
\mbox{Area} & = & 4\pi (r_{+}^2 + a^2) \nonumber \\
\mbox{Surface gravity} & = & \frac{r_{+} - M}{ 2 M r_{+} - Q_{*}^2 } \nonumber
\\
\mbox{Angular velocity} & = & \frac{a}{ r_{+}^2 + a^2 }
\end{eqnarray}

The null tetrad (principle congruences) definitions (the components
refer to $t, r, \theta, \phi$ respectively) : 

\begin{center}
\begin{tabular}{lclclcl}
$ \ell^{\mu}$ & $\equiv$ & $ \frac{1}{\Delta} ( r^2 + a^2 , ~ \Delta, ~
0, ~ a ) ~~~$ &,& ~~
$ n^{\mu}$ & $\equiv$ & $ \frac{1}{2 \eta^2} ( r^2 + a^2 , ~ -\Delta, ~
0, ~ a ) ~~~$ , \\
$ m^{\mu}$ & $\equiv$ & $ \frac{1}{\sqrt{2 \xi} } ( i a sin\theta , ~ 0, 
~ 1, ~ \frac{i}{sin\theta} ) ~~~$ &,& ~~
$ \bar{m}^{\mu}$ & $\equiv$ & $ \frac{1}{\sqrt{2 \bar{\xi}} } ( -i a sin\theta , ~ 0, 
~ 1, ~ \frac{-i}{sin\theta} ) ~~$ . \\
\end{tabular}
\end{center}

These are not well defined on the horizons ($\Delta = 0$) since the
coordinates are singular at the horizon. The $\ell, n$ are both future 
directed and `out-going' and `in-coming' respectively.  Outside the event 
horizon one can make a scaling by $A = 1/\Delta$, and get the $\ell$ to be 
well defined on the horizons. Though $n$ is ill-defined on the horizon, it 
does not matter since we don't not need to use it. This scaling changes some 
of the spin coefficients given by Chandrasekhar. Following are these changed 
values, on horizons, except $\gamma$ which is not needed.

\begin{center}
\begin{tabular}{lcl}
$\kappa = \rho = \sigma = \lambda = \nu = 0 $, & & 
$\epsilon = r - M , ~~~~~~~~ \mu = - \frac{1}{2 \bar{\xi} \eta^2} $, \\
$\pi = \frac{i a sin\theta}{\sqrt{2} \bar{\xi}^2}, ~~~ 
\tau = - \frac{i a sin\theta}{\sqrt{2} \eta^2}$ , & &
$\alpha = \frac{i a sin\theta}{\sqrt{2} \bar{\xi}^2} -
\frac{cot\theta}{2\sqrt{2} \bar{\xi}}, ~~~ 
\beta = \frac{cot\theta}{2 \sqrt{2} \xi}$ , \\
\end{tabular}
\end{center}

Notice that we already have $\alpha + \bar{\beta} = \pi$, and that gauge
covariant derivative of $\pi$ is zero. However $\mu$ is complex. We can make 
a boost transformation to make this real without disturbing the ``gauge 
condition".  Such a transformation parameter, $b$,  satisfies,
$$
\mu - \bar{\mu} + b \pi - \bar{b}\bar{\pi} + 2 \bar{b}\beta - 2
b\bar{\beta} + \delta\bar{b} - \bar{\delta}b = 0 .
$$

In general this gives a one parameter family of solutions. However
regularity on the spherical leaves fixes this uniquely to: 
$$
b = -i \frac{a sin\theta}{2 \sqrt{2} r (r^2 + a^2)} .
$$

It can be explicitly checked that this is consistent with the
commutation relations. \\

Since $\epsilon$ is real, $\alpha - \bar{\beta}$ is unchanged. $\pi$
and $\mu$ change to,

\begin{equation}
\pi^{\prime} = \frac{i a sin\theta}{\sqrt{2}} \left[
\frac{1}{\bar{\xi}^2} ~ + ~ \frac{r - M}{r ( r^2 + a^2 )} \right]
\end{equation}

\begin{equation}
\mu^{\prime} = - \frac{1}{4 r (r^2 + a^2) \eta^2 } \left[ 2 (r^2 - a^2
cos^2 \theta) + \frac{a^2 sin^2 \theta}{r^2 + a^2} \left \{ \frac{M}{r}
\eta^2 + a^2 sin^2 \theta \right \} \right]
\end{equation}

This is manifestly negative definite consistent with marginal trapping. \\

For subsequent computations we note that all quantities we will need will 
be functions only of $\theta$ and hence {\em{effectively}} the  derivatives 
$\delta, \bar{\delta}$ are invariant and can be taken to be,

\begin{equation}
\delta = \frac{1}{\sqrt{2} \xi}\partial_{\theta}, ~~~~~
\bar{\delta} = \frac{1}{\sqrt{2} \bar{\xi}}\partial_{\theta}, ~~~~~
\mbox{effectively}
\end{equation}


The function $f$, that gives the rotational Killing vector, can be
determined by noting that $\bar{\alpha} - \beta$ can be expressed as $-
\delta (\ell n (sin\theta /\xi) )$. Solving $\underline{\delta}^2 f = 0$
gives,

\begin{eqnarray}
f = -\hat{C} cos \theta + \mbox{constant}~~~~~ & , & ~~~~~ \hat{C} ~ = ~ 
-(r^2 + a^2)  \nonumber
\\
C = ~~~ i \bar{\delta} f ~~~~~ & = & ~~~~~  \frac{i\hat{C}}{\sqrt{2} \bar{\xi}} sin \theta
\end{eqnarray}

The constant has been fixed by demanding that the range of the Killing 
parameter is $2\pi$. This is explained below. \\

One can solve the equations determining $\zeta$ and get,

\begin{equation}
\zeta = (\frac{\hat{C}M}{a})\left[ \frac{\eta^2 - 2 r^2}{\eta^4} +
\frac{r^2 - a^2}{( r^2 + a^2 )^2} \right]  ~ + ~ (\frac{\hat{C}Q_{*}^2 r}{a})
\left[ \frac{1}{\eta^4} - \frac{1}{(r^2 + a^2)^2} \right]
\end{equation}

The condition that $\zeta$ must vanish where $C$ vanishes, is used to
fix the constant of integration. \\

It can be checked directly that the invariant vector field $X$ is
already tangential to leaves (has the coefficient of $\ell$ to be zero).
So no further boost transformation is needed. In terms of the
coordinate basis provided by the $t,r,\theta, \phi$, $X$ has non
vanishing components along $t$ and $\phi$ direction, implying that the
adapted azimuthal angle does {\em{not}} coincide with the $\phi$. This
is not surprising since on the event horizon $\phi$ is singular. One can
make a coordinate transformation to the usual non-singular coordinates
$r_{*}, \tilde{\phi}^{+}$ \cite{chandra} and see explicitly that in
terms of these coordinate basis, $X$ has non vanishing component only
along $\tilde{\phi}^{+}$. Demanding that this be equal to 1, fixes the
constant $\hat{C}$ and identifies $\tilde{\phi}^{+}$ as the adapted
azimuthal coordinate. This coordinate transformation of course does not
affect any of the spin coefficients. It does not affect the area 2-form
either.\\

It is straight forward to check that the integral of $\zeta$ over the
sphere precisely equals the angular momentum Komar integral \cite{wald}
for the Kerr-Newman solution {\em evaluated at the horizon}. 

\begin{equation}
J \equiv -\frac{1}{8\pi}\int_{S^2} \zeta ds = 
-\frac{1}{8\pi}\int_{S^2} \zeta (r^2 + a^2) sin\theta d\theta d\phi 
\end{equation}

For comparison we include the Komar integrals corresponding to the
stationary and the axial Killing vectors on a $t = $ constant and $r = $
constant surfaces.

\begin{eqnarray}
M(r) & = & M - Q^2 \left \{ \frac{1}{r} + \frac{a^2}{2 r^3} + \frac{r^2 + a^2}
{2 a r^2}\left( \mbox{arctan} \left( \frac{a}{r} \right) - \frac{a}{r} 
\right) \right \} \nonumber \\
J(r) & = & Ma - Q^2 \left \{ \frac{3 a}{4 r} + \frac{3 a^3}{4 r^3} + 
\frac{(r^2 + a^2)^2} {4 a^2 r^2}\left( \mbox{arctan} 
\left( \frac{a}{r} \right) - \frac{a}{r} \right) \right \} 
\end{eqnarray}

Next, the Ricci scalar of a diagonal metric on $S^2$ is given by,

\begin{equation}
R = -\frac{1}{2}\left[ \partial_{\theta}\left( \frac{\partial_{\theta}
g_{\phi\phi}}{det(g)}\right) + \frac{1}{\det(g)}\partial^2_{\theta}
g_{\phi\phi} \right]
\end{equation}

From our procedure of introducing $\theta,\phi$ coordinates, we have 
$g_{\theta\theta} = -2 C\bar{C}\Phi^2 , g_{\phi\phi} = -2 C\bar{C}$. We
have found above $C$ for the Kerr-Newman family and we also have $R = K
+ \bar{K}$. Thus we get a differential equation for $\Phi$. This can be
solved easily to get, (for both Kerr and Kerr-Newman curvatures)

\begin{equation}
\frac{1}{\Phi^2} = \frac{(r^2 + a^2)^2 sin^2(\theta)}{\eta^4}
\end{equation}

This reduces our standard form of the metric to the usual one for the
Kerr-Newman solution.

\end{document}